\documentclass[reprint,a4paper,aps,prl,longbibliography,amsmath,amsfonts,showpacs,floatfix,nofootinbib]{revtex4-2}
\usepackage{dcolumn}                 
\usepackage{times}
\usepackage{nicefrac}
\usepackage{amssymb}
\usepackage{amsthm}
\usepackage{xcolor}
\usepackage{graphicx}
\newcommand*{\centt}[1]{\multicolumn{1}{c}{#1}}
\newcommand*{\cent}[1]{\multicolumn{1}{c}{$#1$}}
\newcolumntype{x}[1]{D{.}{.}{#1}}
\newcolumntype{y}{>{$}r<{$}}

\newcommand{\header}[1]{\paragraph*{#1.---\hspace{-2ex}}}

\newcommand{\etal}{\textit{et al.}\,}
\allowdisplaybreaks

\begin{document}
\preprint{Version 1.2}

\title{Hyperfine structure of the $2\,^3\!P$ state in $^9$Be and the nuclear quadrupole moment}

\author{Mariusz Puchalski}
\affiliation{Faculty of Chemistry, Adam Mickiewicz University, Uniwersytetu Pozna{\'n}skiego 8, 61-614 Pozna{\'n}, Poland}

\author{Jacek Komasa}
\affiliation{Faculty of Chemistry, Adam Mickiewicz University, Uniwersytetu Pozna{\'n}skiego 8, 61-614 Pozna{\'n}, Poland}

\author{Krzysztof Pachucki}
\affiliation{Faculty of Physics, University of Warsaw, Pasteura 5, 02-093 Warsaw, Poland}

\date{\today}

\begin{abstract}
We have performed accurate calculations of the hyperfine structure of the $2\,^3\!P$ state in the $^9$Be
atom with the help of highly optimized, explicitly correlated functions, accounting for the leading finite nuclear
mass, radiative, nuclear structure and relativistic effects. By comparison with measurements, we
have determined the $^9$Be nuclear quadrupole moment to be $Q_{\rm N} = 0.05350(14)$ barns,
which is not only the most accurate result, but also disagrees with previous determinations.
\end{abstract}

\maketitle

The electric quadrupole moment $Q_{\rm N}$ of nuclei is a measure of the deformation 
of the nuclear charge distribution with respect to the spherical symmetry. 
This deformation comes from a strong dependence of nuclear forces on the orientation of the nucleon spin
and is present in atomic nuclei with a spin equal to or greater than one, as it is for $\,^9$Be with $I = 3/2$. 
In principle, $Q_{\rm N}$ could be determined from the nuclear structure theory, but 
such calculations with controlled accuracy have not yet been feasible. 
Only very recently and only for such a simple nucleus as the deuteron, 
theoretical predictions based on the chiral effective field theory have reached the uncertainty of 1\% \cite{Filin:20b}, 
and this result agreed with the more precise value from the rotational spectroscopy of the deuterium hydride (HD) molecule \cite{Puchalski:20}. 

Originally,
$Q_{\rm N}$ for many nuclei was determined by the electron scattering of the nuclei,
for example  for $^7$Li \cite{Voelk:91}. However, for $^9$Be the available experimental data 
are not only limited in accuracy but also differ noticeably from each other, indicating that the
scattering results are not always conclusive \cite{Vinciguerra:69,Bernheim:69}. 
Nowadays, reliable $Q_{\rm N}$ values can be derived from a combination of theoretical 
and atomic or molecular spectroscopic data. This has been realized for LiH, LiF, 
and LiCl molecules \cite{Urban:90}, leading to a $Q_\mathrm{N}(^7\mathrm{Li})$ consistent 
with the nuclear scattering value \cite{Voelk:91}. Although appropriate theoretical 
calculations were performed for BeH$^+$ \cite{Diercksen:89,Borin:92} as well as 
for the $^{7,9}$Be$^-$ anion \cite{Nemouchi:03} years ago, there have been no corresponding measurements reported so far.
Therefore, the method of choice for finding $Q_\mathrm{N}(^9\mathrm{Be})$ 
is atomic spectroscopy. 

In the accurate determination of $Q_{\rm N}$ from atomic spectroscopy, it is important to understand 
the electron-nucleus interaction at the fundamental level. 
Recent advances in measurements of electronic \cite{CODATA:18} and muonic atoms \cite{Krauth:21},
together with progress in the theoretical description of atomic spectra \cite{Pachucki:18},
indicate the importance of the nuclear structure  effects. Therefore, accurate knowledge of the nuclear quadrupole moment 
will give access to details of the electron-nucleus interactions 
which, so far, have not been visible, like for example the nuclear quadrupole polarizability \cite{Friar:05}. 

The ground state $2\,^1\!S_0$ of Be is fully symmetric; thus, 
the nuclear quadrupole moment does not lead here to any observable effect. In contrast, 
in the lowest excited $2\,^3\!P$ level, which is metastable, the hyperfine interactions cause 
level splitting, and this splitting can be accurately measured \cite{Blachman:67}. 
So, it is atomic structure theory that must interpret the hyperfine structure in terms of the magnetic dipole and the
electric quadrupole moments of the nucleus. Still, the same hyperfine interactions that
lead to the hyperfine splitting also cause the hyperfine mixing of different $2\,^3\!P_J$ levels.
This mixing had been accounted for in Ref.~\onlinecite{Blachman:67} but in a very
simplified way. Although the accuracy of these approximations has been questioned
\citep{Diercksen:89}, consecutive atomic structure calculations, using modern and sophisticated
approaches, have relied on these original inaccurate calculations of the hyperfine mixing. 
An exception is the work by Beloy \etal \cite{Beloy:08}, in which this mixing was calculated 
in the second order of perturbation theory using the relativistic CI+MBPT approach, 
but its numerical precision turned out to be insufficient to obtain $Q_{\rm N}$ 
with a competitive accuracy. 
This hyperfine mixing is not specific just to Be atoms, it is certainly present in all the other atomic systems.
Therefore, the determinations of quadrupole moments for many other nuclei might have lower accuracy than anticipated,
what should be verified, as we have done here for the $^9$Be nucleus.

In this work, we employ well-optimized explicitly correlated Gaussian wave functions to have 
good control over the numerical accuracy. With these wave functions we perform calculations 
of the $^9$Be hyperfine structure, with a complete treatment of the hyperfine mixing of different 
$2\,^3\!P_J$ levels. Moreover, we fully account for the leading radiative (QED), finite nuclear mass, 
and nuclear structure effects, while dealing with relativistic and higher order corrections approximately.  
Our result for the electric quadrupole moment $Q_{\rm N}(^9{\rm Be})$ show that all of its previous 
determinations were not as accurate as claimed, due to a very approximate treatment of the hyperfine 
mixing and neglect of the radiative (QED) corrections.

\header{Fine and hyperfine structure Hamiltonian}
The most accurate approach for light atomic systems  solves at first the nonrelativistic Hamiltonian
\begin{eqnarray}
H &=& \frac{\vec p^{\,2}_\mathrm{N}}{2\,m_\mathrm{N}} + \sum_a \frac{\vec p^{\,2}_a}{2} - \sum_a \frac{Z}{r_a} + \sum_{a<b} \frac{1}{r_{ab}},
\end{eqnarray}
and next includes relativistic effects as a perturbation.
Relevant relativistic effects are split into the fine structure Hamiltonian
\begin{align}
H_{\rm fs} =&\ Q_1 + Q_2 \label{EHfs}\\
Q_1 =&\ \sum_a\,\frac{Z\,\alpha}{4}\,\vec{\sigma}_a \cdot \Big[ \frac{(g-1)}{m^2} 
\frac{\vec{r}_a}{r_a^3}\times\vec{p}_a  - \frac{g}{m\,m_{\rm N}}\,\frac{\vec r_a}{r_a^3}\times\vec p_{\rm N} \Big]  \nonumber \\
&\hspace{-0.5cm} +\sum_{a\neq b}\, \frac{\alpha}{4\,m^2}\,
\vec{\sigma}_a \cdot \Big[g\,\frac{\vec{r}_{ab}}{r_{ab}^3} \times\vec{p}_b-(g-1)\,\frac{\vec{r}_{ab}}{r_{ab}^3} \times\vec p_a\Big] \\
Q_2=&\ - \sum_{a<b} \frac{3\,\alpha\,g^2}{16\,m^2}\,\sigma_a^i\,\sigma_b^j\,\biggl(
\frac{r_{ab}^i\,r_{ab}^j}{r_{ab}^5} - \frac{\delta^{ij}}{3\,r_{ab}^3}\biggr) 
\end{align}
and the hyperfine structure Hamiltonian
\begin{align}\label{EHhfs}
H_{\rm hfs} &=\vec I\cdot\vec Q + \frac{3\,I^i\,I^j}{I\,(2\,I-1)}\,\frac{Q^{ij}}{6}\\
\vec Q &=\vec Q_1+\vec Q_2 + \vec Q_3  \\
\vec Q_1 &=\sum_a
\frac{1}{3}\,\frac{Z\,\alpha\,g\,g_{\rm N}}{m\,m_{\rm N}}\,
\vec \sigma_a\,\pi\,\delta^3(r_a) \\ 
\vec Q_2 &=\sum_a
\frac{Z\,\alpha\,g_\mathrm{N}}{2\,m\,m_\mathrm{N}}\,\frac{\vec r_a}{r_a^3}\times\vec p_a
-\frac{Z\,\alpha\,(g_\mathrm{N}-1)}
{2\,m_\mathrm{N}^2}\,\frac{\vec r_a}{r_a^3}\times\vec p_{\rm N} \\ 
Q_3^j &=\sum_a
\frac{Z\,\alpha\,g\,g_\mathrm{N}}{8\,m\,m_\mathrm{N}}\,\sigma_a^i\,
\biggl(\frac{3\,r_a^i\,r_a^j}{r_a^5} - \frac{\delta^{ij}}{r_a^3}\biggr) \\ 
Q^{ij} &=-\sum_a
\alpha\,Q_\mathrm{N}\,
\biggl(\frac{3\,r_a^i\,r_a^j}{r_a^5} - \frac{\delta^{ij}}{r_a^3}\biggr)\,, \label{EQij}
\end{align}
where $\vec p_{\rm N} = -\sum_a \vec p_a$ is the nuclear momentum, $\vec{I}$ is the nuclear spin, $g$ is the free-electron g-factor, 
$Q_\mathrm{N}$ is the electric quadrupole moment of the nucleus, 
and $g_{\rm N}$ is the nuclear g-factor 
\begin{equation}
g_{\rm N} = \frac{m_{\rm N}}{Z\,m_{\rm p}}\,\frac{\mu}{\mu_{\rm N}}\,\frac{1}{I}.
\end{equation}
Leading QED effects are included in the free electron g-factor, while higher order corrections 
are accounted for as described later on.

Because the hyperfine splittings are only about 100 times smaller than the fine splitting,  
one cannot assume that hyperfine levels have a definite angular momentum $\vec J= \vec L + \vec S$,
but only definite $\vec L$ and $\vec S$. It is convenient therefore to extend the original formulation of the hyperfine splitting
by Hibbert in \cite{Hibbert:75} and represent  the fine and the hyperfine structure of the $2\,^3\!P$ state in terms of an effective Hamiltonian,
instead of expectation values only
\begin{align}\label{EHeff}
H_\mathrm{eff}=&\ c_1\,\vec{L}\cdot\vec{S} +c_2\,(L^iL^j)^{(2)}(S^iS^j)^{(2)}
\nonumber \\ & \
+a_1\,\vec{I}\cdot\vec{S}+a_2\,\vec{I}\cdot\vec{L}+a_3\,(L^iL^j)^{(2)}S^iI^j 
\nonumber \\ & \
+\frac{b}{6}\,\frac{3\,(I^i\,I^j)^{(2)}}{I\,(2\,I-1)}\,\frac{3\,(L^iL^j)^{(2)}}{L\,(2\,L-1)},
\end{align}
where the coefficients $a_1, a_2, a_3, b, c_1,c_2$ are independent of $J$,
but are specific to the particular state. These coefficients can be obtained, for example,
from the matrix elements of $H_{\rm fs}$ and $H_{\rm hfs}$ in the 
decoupled $|M_L,M_S\rangle$ or $|J, M_J\rangle$ basis. Once these coefficients are known, 
the above Hamiltonian can be diagonalized yielding the hyperfine levels.
Alternatively, the hyperfine structure can be represented in terms of $J$-dependent
$A_J$ and $B_J$ coefficients
\begin{align}\label{EHhfseff}
H_{\rm hfs,\, eff} = A_J \vec I\cdot \vec J + \frac{B_J}{6}\,\frac{3\,(I^i\,I^j)^{(2)}}{I(2\,I-1)}\,\frac{3\,(J^i\,J^j)^{(2)}}{J(2\,J-1)},
\end{align}
which are conventionally used to represent the measured values, while the fine structure 
is given in terms of differences in the centroid energies.

If the atomic levels had a definite value $J$, the fine structure would be given by the expectation value of $H_{\rm eff}$, namely
\begin{align}
E_{\rm fs}(J) 
       =&\ \left\{\begin{array}{lr}
       c_1+c_2/6 &\mbox{\rm for} \;J=2\\
       -c_1-5\,c_2/6 &\mbox{\rm for} \;J=1 \\
       -2\,c_1+5\,c_2/3 &\mbox{\rm for}\; J=0
       \end{array}\right.,
\end{align}
and the hyperfine structure  by 
\begin{align}
A_J =&\  \left\{\begin{array}{lr} 
a_1/2+ a_2/2 + a_3/6&\mbox{\rm for} \;J=2\\
a_1/2+ a_2/2 - 5\,a_3/6&\mbox{\rm for} \;J=1
\end{array}\right.
\,,\nonumber\\
B_J =&\  \left\{\begin{array}{lr}
b&\mbox{\rm for} \;J=2\\
-b/2&\mbox{\rm for} \;J=1
\end{array}\right.\,.
\end{align}
Because, in general, one cannot assume that the levels have a definite value of $J$, 
the effective hyperfine Hamiltonian in Eq.~(\ref{EHeff}) has to be diagonalized numerically. 
Nonetheless, the hyperfine structure can still be represented in terms of $A_J$ and $B_J$
coefficients, and we use them for comparison of theoretical predictions with experimental results
and for the determination of the nuclear quadrupole moment $Q_{\rm N}$.

\header{Relativistic, radiative, and finite nuclear size corrections}
The hyperfine Hamiltonian in Eq.~(\ref{EHhfs}) represents the leading hyperfine interactions, 
but there are also many small corrections which are often overlooked in literature. 
These corrections contain terms with higher powers of the fine structure constant $\alpha$.
Because most of them are proportional to the Fermi contact interaction, 
i.e. to $\vec{I}\cdot\vec{Q}_1$ in Eq.~(\ref{EHhfs}), we account for them
in terms of the following factor
\begin{align}
\tilde a_1 = a_1(1+\epsilon).
\end{align}
Below, we briefly describe contributions included in the $\epsilon$ term.

The $\mathcal{O}(\alpha)$ correction is analogous to that in hydrogenic systems \cite{Eides:01} 
and consists of two parts. The first part is due to the nuclear recoil
\begin{align}
H^{(5)}_{\rm rec} =&\  \biggl[-\frac{3\,Z\,\alpha}{\pi}\,\frac{m}{m_{\rm N}}\,\ln\Bigl(\frac{m_{\rm N}}{m}\Bigr)\biggr]
\vec I\cdot \sum_a \frac{2}{3}\,\frac{Z\,\alpha\,g_{\rm N}}{m\,m_{\rm N}}\,\vec \sigma_a\,\pi\,\delta^3(r_a) 
\end{align}
and numerically is very small, almost negligible. The recoil contribution to $\epsilon$ 
amounts to $\epsilon_{\rm rec} = -0.000\,011$. The second part of the $\mathcal{O}(\alpha)$ correction is due to 
the finite nuclear size and the nuclear polarizability, and is given by \cite{Eides:01, Puchalski:09}
\begin{align}
H^{(5)}_{\rm fs} =&\  \bigl[-2\,Z\,\alpha\,m\,r_Z\bigr]
\vec I\cdot \sum_a \frac{2}{3}\,\frac{Z\,\alpha\,g_{\rm N}}{m\,m_{\rm N}}\,\vec \sigma_a\,\pi\,\delta^3(r_a) ,
\end{align}
where $r_Z$ is a kind of effective nuclear radius called the Zemach radius. 
Disregarding the inelastic effects, this radius can be written down in terms of 
the electric charge $\rho_E$ and magnetic-moment $\rho_M$ densities as
\begin{align}
r_Z = \int d^3 r\,d^3r'\,\rho_E(r)\,\rho_M(r')\,|\vec r-\vec r'|.
\end{align}
Nevertheless, the inelastic, i.e. polarizability, corrections can be significant, but 
because they are very difficult to calculate, they are usually neglected. 
In this work we account for possible inelastic effects
by employing  $r_Z$ from a comparison of very accurate
calculations of hfs in $^9$Be$^+$ with the experimental value, namely  $r_Z = 4.07(5)\;{\rm fm}$~\cite{Puchalski:09}. 
Because this correction is also proportional to the contact Fermi interaction, 
we represent it in terms of $\epsilon_{\rm fs} = -0.000\,615$.

Next, there are radiative and relativistic corrections of the relative order $\mathcal{O}(\alpha^2)$.
The radiative correction, beyond that included by the free electron g-factor, is \cite{Eides:01}
\begin{align}
H^{(6)}_{\rm rad} = Z\,\alpha^2\biggl(\ln 2-\frac{5}{2}\biggr)
\vec I\cdot \sum_a \frac{2}{3}\frac{Z\,\alpha\,g_{\rm N}}{m\, m_{\rm N}}\,\vec \sigma_a\,\pi\,\delta^3(r_a)
\end{align}
and the corresponding $\epsilon$ factor is $\epsilon_{\rm rad} = -0.000\,384$.
The $\mathcal{O}(\alpha^2)$ relativistic and higher order corrections are much more complicated. 
They have been calculated for the ground state of $\,^9$Be$^+$ \cite{Puchalski:09}.
Here we take this result and assume that it is proportional to the Fermi contact interaction, and obtain
$\epsilon_{\rm rel} = 0.001\,664$. This is the only approximation we assume in this work, 
and as a consequence we neglect the mixing of the $2\,^3\!P_1$ state with the nearby lying $2\,^1\!P_1$.
Exactly for this reason we will use only the hyperfine splitting of the $2\,^3\!P_2$ state, which does not mix with $2\,^1\!P_1$
for the determination of $Q_{\rm N}$. The resulting total correction is 
\begin{align}
\epsilon =  \epsilon_{\rm rec} + \epsilon_{\rm fs} + \epsilon_{\rm rad} + \epsilon_{\rm rel} = 0.654\cdot10^{-3}.
\end{align}
Some previous works present these multiplicative correction for all individual hyperfine contributions,
but in our opinion this cannot be fully correct because higher order relativistic corrections
may include additional terms, beyond that in $H_{\rm eff}$ in Eq.~(\ref{EHeff}). These corrections
are expected to be much smaller than the experimental uncertainty for the $B_J$ coefficient,
and therefore are neglected.

\header{ECG wave function and expectation values}
Let us now move to the calculations of the $a_i, b$, and $c_i$ coefficients.
To obtain sufficiently high accuracy for these parameters, we express the four-electron 
atomic wave function as a linear combination of properly symmetrized explicitly correlated 
Gaussian (ECG) functions,
\begin{equation}
\Psi\left(\{ \vec r_a\}\right) = \sum_{n=1}^K t_n \,{\cal A}\left[\,\phi_n\left(\{ \vec r_a\}\right)\,\chi_{\{a\}}\right] \,,
\end{equation}
where $t_n$ are linear coefficients, $\cal A$ is the antisymmetrization operator over electronic indices,
and $\{a\}$ and $\{\vec r_a\}$ denote the sequence of electron indices and coordinates, respectively.
The electronic $P$ symmetry of the states was enforced using the following spatial functions
\begin{eqnarray}
\phi^i(\{ \vec r_a\}) &=&  r^i_a\,\exp \big[-\sum_b w_b \,r^2_b -\sum_{c<d} u_{cd} \,r^2_{cd} \big] \label{phiP} 
\end{eqnarray}
with $w_b$ and $u_{cd}$ being nonlinear variational parameters,
while the spin functions corresponding to different spin $S=1$ projections were, for $\{a\} = (1,2,3,4)$:
\begin{eqnarray}
\chi^{1-1}_{\{a\}} &=& \frac{1}{\sqrt{2}} (\alpha_1\,\beta_2-\beta_1\,\alpha_2)\,\beta_3\,\beta_4\,, \\
\chi^{10}_{\{a\}} &=& \frac{1}{2} (\alpha_1\,\beta_2-\beta_1\,\alpha_2) (\alpha_3\,\beta_4-\beta_4\,\alpha_3), \\
\chi^{11}_{\{a\}} &=& \frac{1}{\sqrt{2}} (\alpha_1\,\beta_2-\beta_1\,\alpha_2)\,\alpha_3\,\alpha_4\,,
\end{eqnarray}
where $\alpha_i$ and $\beta_i$ are the one-electron spin-up and spin-down functions.
To control the uncertainty of our results we performed the calculations with several basis sets successively increasing their size. The nonlinear parameters were optimized variationally
with respect to the nonrelativistic energy $E$ until the energy reached stability in a desired number of digits. From the analysis of convergence we obtained the extrapolated nonrelativistic 
energies and mean values used in Table I.
The largest wave functions optimized variationally were composed of 6144 terms, leading to the nonrelativistic energy in agreement with the result obtained by
Kedziorski et al. \cite{Kedziorski:20}.

\begin{table*}[!hbt]
\renewcommand{\arraystretch}{1.0}
\caption{Convergence of the nonrelativistic energy and theoretical fine and hyperfine structure
parameters for the $2\,^3\!P$ state of $^9$Be (in MHz). Mass $m_\mathrm{N}= 9.012\,183\,07(8)$ u 
\cite{Wang:17} and magnetic moment $\mu/\mu_{\rm N} = -1.177\,432(3)$ \cite{Stone:15} were used
for the $^9$Be nucleus.} 
\label{Tabcparms}
\begin{ruledtabular}
\begin{tabular}{rx{3.14}x{6.4}x{5.4}x{4.7}x{4.7}x{3.7}x{3.7}}
\rule[-1.0ex]{0pt}{2ex}$K$& \cent{E/\mathrm{a.u.}} & \cent{c_1}& \cent{c_2} & \cent{a_1} & \cent{a_2} & \cent{a_3}& \cent{b/Q_\mathrm{N}\,{\rm (MHz/barn)}} \\
\hline
1536     & -14.566\,340\,608\,5 & 32\,986.385 & 5\,399.032  & -231.116\,70 & -22.698\,781   & 14.788\,494  & 27.150\,476 \\
2048     & -14.566\,341\,144\,7 & 32\,987.849 & 5\,399.186  & -231.126\,66 & -22.698\,923   & 14.788\,389  & 27.149\,896 \\
3072     & -14.566\,341\,380\,0 & 32\,989.401 & 5\,399.240  & -231.130\,99 & -22.698\,986   & 14.788\,467  & 27.149\,112 \\
4096     & -14.566\,341\,441\,5 & 32\,989.854 & 5\,399.190  & -231.129\,61 & -22.699\,124   & 14.788\,579  & 27.148\,954 \\
6144     & -14.566\,341\,466\,0 & 32\,990.027 & 5\,399.172  & -231.128\,99 & -22.699\,186   & 14.788\,630  & 27.148\,898 \\
$\infty$ & -14.566\,341\,474(8) & 32\,990.2(2)& 5\,399.15(2)& -231.128\,4(6)& -22.699\,24(5) & 14.788\,68(5)& 27.148\,87(3)
\end{tabular}
\end{ruledtabular}
\end{table*}

The optimized nonrelativistic wave functions were subsequently employed to evaluate 
the fine and hyperfine parameters from the expectation values of the corresponding 
quantum-mechanical operators~(\ref{EHfs})-(\ref{EQij}). 
For this purpose we chose $J=2$, $L=1$, and for an arbitrary $M$
\begin{align}
c_1&= \langle J,M| Q_1 |J,M\rangle,\\
c_2&= 6\,\langle J,M| Q_2 |J,M\rangle ,\\
a_1&= \langle J,M| \vec Q_1\cdot\vec J |J,M\rangle/3,\label{Ea1}\\
a_2&= \langle J,M| \vec Q_2\cdot\vec J |J,M\rangle/3 ,\\
a_3&= \langle J,M| \vec Q_3\cdot\vec J |J,M\rangle,\\
b  &= \langle L,M| Q^{ij}\,L^i\,L^j |L,M\rangle/5.
\end{align}
The numerical values of these parameters and their convergence with increasing size 
of the ECG basis are shown in Table~\ref{Tabcparms}. 
Their relative accuracy can be estimated collectively as better than $6\cdot 10^{-6}$.

\header{Fine and hyperfine structure} 
The hyperfine transition frequencies $\nu_{\!J}(F;F+1)$ can be expressed in terms 
of the $A_J$ and $B_J$ parameters and  vice versa. 
For this, one writes
\begin{align}
\langle H_{\rm hfs, eff}\rangle_F =&\ A_J\,A_{IJF} + B_J\,B_{IJF} + C_J\,C_{IJF}
\end{align}
where 
\begin{align}
A_{IJF} &=\frac{1}{2}K \\
B_{IJF} &=\frac{3/4\, K (K + 1) - I (I + 1) J (J + 1)}{2 I (2 I - 1) J (2 J - 1)}
\end{align}
with $K=F (F + 1) - I (I + 1) - J (J + 1)$ and the total angular momentum
$\vec{F}=\vec{S}+\vec{L}+\vec{I}$.
The octupole term $C_{IJF}$ is given e.g. by Schwartz \cite{Schwartz:55} and Jaccarino \cite{Jaccarino:54}.

For $J=1$ we arrive at
\begin{align}
A_1&=-\frac{1}{6}\,\nu_1\!\left(\frac{1}{2};\frac{3}{2}\right)-\frac{3}{10}\,\nu_1\!\left(\frac{3}{2};\frac{5}{2}\right)\label{EA1fr}\\
B_1&=\frac{1}{3}\,
   \nu_1\!\left(\frac{1}{2};\frac{3}{2}\right)-\frac{1}{5}\,\nu_1\!\left(\frac{3}{2};\frac{5}{2}\right)\,, \label{EB1fr}
\end{align}
and for $J=2$
\begin{align}
A_2&=-\frac{3}{50}\,\nu_2\!\left(\frac{1}{2};\frac{3}{2}\right)-\frac{7}{50}\,\nu_2\!\left(\frac{3}{2};\frac{5}{2}\right)-\frac{4}{25}\,\nu_2\!\left(\frac{5}{2};\frac{7}{2}\right) \label{EA2fr}\\
B_2&=\frac{2}{5}\,
   \nu_2\!\left(\frac{1}{2};\frac{3}{2}\right)+\frac{2}{5}\,\nu_2\!\left(\frac{3}{2};\frac{5}{2}\right)-\frac{16}{35}\,\nu_2\!\left(\frac{5}{2};\frac{7}{2}\right) \label{EB2fr}\\
C_2&=-\frac{1}{50}\,
   \nu_2\!\left(\frac{1}{2};\frac{3}{2}\right)+\frac{1}{50}\,\nu_2\!\left(\frac{3}{2};\frac{5}{2}\right)-\frac{1}{175}\,\nu_2\!\left(\frac{5}{2};\frac{7}{2}\right) .\label{EC2fr}
\end{align}
\begin{table}[!hbt]
\renewcommand{\arraystretch}{1.0}
\caption{$A_J$, $B_J$, and $C_J$ parameters determined using Eqs.~(\ref{EA1fr})-(\ref{EC2fr})
from experimental \cite{Blachman:67} and theoretical 
hfs frequencies $\nu_{\!J}$ (in MHz). 
The theoretical $B_2$ is by definition equal to the experimental one, which serves for determination of $b$ in Eq. (\ref{Eb}).}
\label{TABC}
\begin{ruledtabular}
\begin{tabular}{lx{3.8}x{4.5}x{3.8}}
& \centt{Experimental} & \centt{Theoretical} & \centt{Difference} \\
\hline
$A_1$ & -140.1564(8)  & -140.1091  & -0.0473(8)  \\   
$B_1$ &   -3.2363(7)  &   -3.2337  & -0.0026(7)  \\[2ex]
$A_2$ & -124.6165(5)  & -124.6073  & -0.0092(5)  \\
$B_2$ &    0.7800(21) &    0.7800  &  0.0        \\
$C_2$ &   -0.00030(9) &   -0.00032 &  0.00002(9) 
\end{tabular}
\end{ruledtabular}
\end{table}
The above listed parameters~(\ref{EA1fr})-(\ref{EC2fr}) were evaluated using both experimental 
and theoretical transition frequencies $\nu_{\!J}(F;F+1)$, and their numerical values 
are presented in Table~\ref{TABC}. The experimental values of $\nu_{\!J}$ (in MHz) were 
taken from Ref.~\onlinecite{Blachman:67}, assuming their uncertainties are statistical and uncorrelated.
The resulting experimental values for $A_1$ and $B_1$ are the same as in  Ref.~\onlinecite{Blachman:67},
while for $A_2$ and $B_2$ are twice more accurate due to inclusion of the $C_2$ coefficient. 
On the other hand, theoretical values were found by diagonalization of the effective Hamiltonian of Eq.~(\ref{EHeff}).
Most importantly, the parameter $b$ was fixed by matching theoretical and experimental $B_2$, which holds for 
\begin{align}
b=1.4525(40)\;{\rm  MHz}, \label{Eb}
\end{align}
where the uncertainty originates from that of the experimental one in $B_2$.

The differences between experimental and theoretical values, shown in Tab.~\ref{TABC}, 
are not discrepancies but result from neglected higher order relativistic and QED effects,
which are beyond those proportional to the Fermi contact interaction.
For the $A_2$ coefficient these effects are of relative order $0.07\cdot10^{-3}$, 
while for $A_1$ they are $0.33\cdot10^{-3}$. They are larger for $A_1$
due to hyperfine mixing of $2\,^3\!P_1$ with the nearby  $2\,^1\!P_1$ state, which we have not 
taken into account in our calculation.
Moreover, the relative difference for the $B_1$ coefficient
is $\delta B_1/B_1 = 0.80\cdot10^{-3}$, which most probably is also due to this mixing.
For this reason, we have chosen $B_2$ and not $B_1$ for the determination of the $b$ coefficient.

The centroids of the calculated hyperfine energy levels are shown in Tab.~\ref{Tfs} in the column
`fs levels' for all $J$ values of the $2\,^3\!P$ state. This table contains also experimental \citep{Blachman:67} 
and calculated fine-structure transition frequencies as well as their difference.
\begin{table}[!hbt]
\renewcommand{\arraystretch}{1.0}
\caption{Theoretical fine structure (fs) energy levels of the $2\,^3\!P$ state of $^9$Be, 
and the comparison of experimental \cite{Blachman:67} and theoretical fs transition frequencies
(in GHz).}
\label{Tfs}
\begin{ruledtabular}
\begin{tabular}{lx{4.4}x{2.5}x{2.3}x{2.5}}
\rule[-1.0ex]{0pt}{2ex}$J$ & \centt{fs levels} & \cent{\nu^\mathrm{exp}(J;J+1)}  & \cent{\nu^\mathrm{the}(J;J+1)} & \centt{Diff.} \\
\hline
2   &  33.890 \\
1   & -37.489 & 71.86(24) & 71.379 &  0.48(24) \\
0   & -56.987 & 19.41(19) & 19.498 & -0.09(19) \\
\end{tabular}
\end{ruledtabular}
\end{table}
This difference is much larger than the estimate of relativistic corrections, which is $\mathcal{O}((Z\,\alpha)^2)$.
Therefore, we presume, that these differences are due to inaccurate experimental results, because
these fine splittings have not been directly measured
but determined from the magnetic field dependence of the hfs splittings \citep{Blachman:67}.

\header{Determination of $Q_N(^9{\rm Be})$}
Having calculated the value of $b/Q_\mathrm{N}=27.148\,87(3)$ MHz/barn, 
see Tab.~\ref{Tabcparms}, and fixed $b$ by matching theoretical and experimental values
of $B_2$, we obtain the electric nuclear quadrupole moment $Q_\mathrm{N}$ of $\,^9$Be
\begin{align}
Q_{\rm N}&=0.053\,50(14)\text{ barn}\,.
\end{align}
The uncertainty in $Q_\mathrm{N}$ comes from the experimental error in the $\nu_2$ hyperfine frequencies.
In this context, the uncertainties from the neglected higher order relativistic and QED effects 
in the $b$ parameter, as well as the numerical uncertainties, are much smaller
and are not shown.

\begin{figure}[!b]
\includegraphics[scale=0.2]{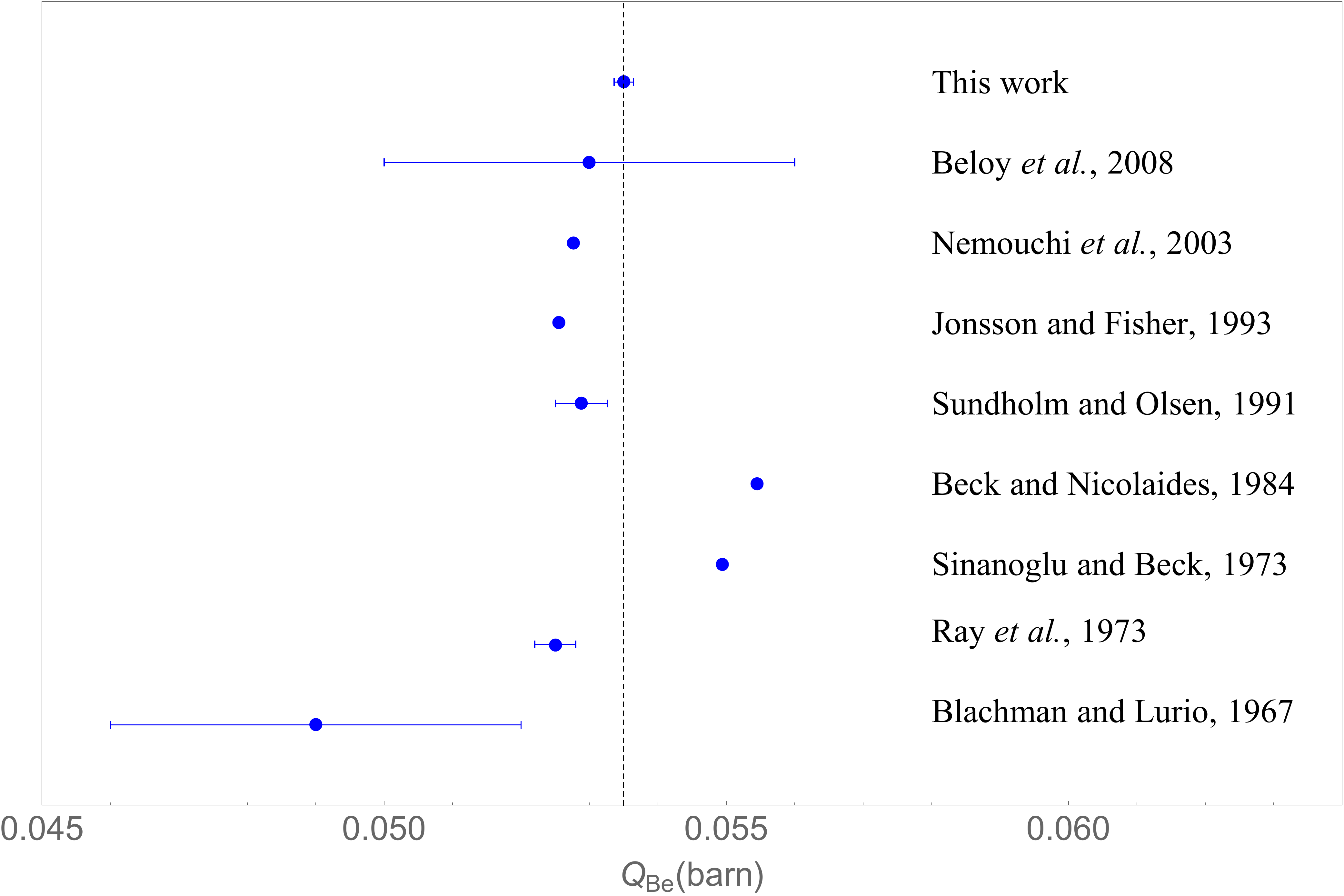}
\caption{Comparison of the electric quadrupole moment (in barns) of the $^9$Be nucleus obtained
using atomic structure calculations.}
\label{Fliterature}
\end{figure}

Among the literature results listed in Tab.~\ref{Tliterature} and depicted in Fig.~\ref{Fliterature},
only the $Q_\mathrm{N}$ reported by Beloy \etal\cite{Beloy:08}, thanks to its large
uncertainty, agrees with our quadrupole moment. 
The lack of agreement with the other previous studies 
indicates that it is very difficult to estimate theoretical  uncertainties
using well-established, atomic structure methods such as MCHF or CI+MBPT.
\begin{table}[!hbt]
\caption{Comparison of the electric quadrupole moment (in barns) of the $^9$Be nucleus obtained
using atomic structure calculations.}
\label{Tliterature}
\begin{ruledtabular}
\begin{tabular}{lx{5.11}}
\rule[-1.0ex]{0pt}{2ex}Source & \cent{Q_\mathrm{N}(^9{\rm Be})} \\
\hline
Blachman and Lurio, 1967 \cite{Blachman:67}  &  0.049(3) \\
Ray \etal, 1973          \cite{Ray:73}       & 0.052\,5(3) \\
Sinanoǧlu and Beck, 1973 \cite{Sinanoglu:73} & 0.054\,94 \\
Beck and Nicolaides, 1984 \cite{Beck:84}     & 0.055\,45 \\
Sundholm and Olsen, 1991 \cite{Sundholm:91}  & 0.052\,88(38) \\
J\"onsson and Fisher, 1993 \cite{Jonsson:93} & 0.052\,56 \\
Nemouchi {\em et al.}, 2003 \cite{Nemouchi:03}     & 0.052\,77    \\
Beloy \etal, 2008 \cite{Beloy:08}            & 0.053(3) \\
This work                                    & 0.053\,50(14) \\
\end{tabular}
\end{ruledtabular}
\end{table}

\header{Conclusions}
We have determined the electric quadrupole moment $Q_{\rm N}(^9{\rm Be})$ with significantly
higher accuracy and in disagreement with previous values. 
The improvement achieved in this work has several independent sources. 
The first one is the recalculation of  the $A_2$, $B_2$, and $C_2$ coefficients 
from experimental hyperfine splitting. 
More precisely, including $C_2$ in the effective Hamiltonian $H_{\rm hfs,eff}$ in Eq.~(\ref{EHhfseff}) 
enabled us to decrease the experimental uncertainties by a factor of 2. 
The second improvement is due to the inclusion of the hyperfine mixing 
by exact diagonalization of the effective fine- and hyperfine-structure Hamiltonian of Eq.~(\ref{EHeff}). 
The third improvement comes from the very accurate calculation of the expectation values
of the fine- and hyperfine-structure operators using explicitly correlated functions
allowing for the complete electron correlations.
The fourth one is due to the exact inclusion of the finite nuclear mass in the fine and hyperfine 
interaction in Eqs.~(\ref{EHfs}) and (\ref{EHhfs}). Finally, the fifth source of improvement is due to
accounting for the relativistic, radiative and nuclear structure corrections by appropriate rescaling of the $a_1$ parameter.

Regarding our theoretical uncertainties, they come exclusively from neglected higher order
relativistic and QED contributions, in particular those due to hyperfine mixing 
of the $2\,^3\!P_1$ and $2\,^1\!P_1$ states. The calculation of the complete $\mathcal{O}(\alpha^2)$ correction
is possible but technically difficult. It has been performed for Be$^+$ in \cite{Puchalski:14} 
due to the availability of the exponentially correlated basis functions there.
Nevertheless, if a complete $\mathcal{O}(\alpha^2)$ correction is known, one can use both $B_1$ and $B_2$ parameters
to obtain an even more accurate nuclear quadrupole moment of $^9$Be. 
              
\header{Acknowledgments}
This research was supported by National Science Center (Poland) Grant No. 2014/15/B/ST4/05022 
as well as by a computing grant from Pozna\'n Supercomputing and Networking Center and by PL-Grid Infrastructure.

%

\end{document}